\documentclass[conference]{IEEEtran}

\ifCLASSINFOpdf
\else
\fi
\usepackage{graphicx}
\usepackage{setspace}
\usepackage{amsmath,empheq}
\usepackage{amssymb}
\usepackage{amsthm}
\usepackage{caption}
\usepackage{subcaption}
\usepackage{bm} 
\usepackage{cite}
\usepackage{float}
\usepackage[usenames,dvipsnames]{pstricks}
\usepackage{epsfig}
\usepackage{pst-grad} % For gradients
\usepackage{pst-plot} % For axes
\usepackage{tabularx}
\RequirePackage{flushend}
\usepackage{amsmath}
\usepackage[T1]{fontenc}
\newcommand{\lb} {\left}
\newcommand{\rb} {\right}
\newcommand{\nn} {\nonumber}
\usepackage{xcolor}
\usepackage{lipsum}
\usepackage{comment}

\allowdisplaybreaks

\allowdisplaybreaks
\usepackage[utf8]{inputenc}

 \IEEEaftertitletext{\vspace{-2.5\baselineskip}}
 \usepackage{svg}

\usepackage{algpseudocode}
\usepackage{algorithm}
\usepackage{xcolor}
\algnewcommand\algorithmicforeach{\textbf{for each}}
\algdef{S}[FOR]{ForEach}[1]{\algorithmicforeach\ #1\ \algorithmicdo} 

\begin{document}
% \bstctlcite{IEEEexample:BSTcontrol}
\bstctlcite{IEEEexample:BSTcontrol}
\title{Secure Energy Efficient Wireless Transmission: A Finite v/s Infinite-Horizon RL Solution}

  \author{
    \IEEEauthorblockN{Shalini Tripathi\IEEEauthorrefmark{1}, 
    Ankur Bansal\IEEEauthorrefmark{2},
    Holger Claussen\IEEEauthorrefmark{3},
    Lester Ho\IEEEauthorrefmark{4},
    and Chinmoy~Kundu\IEEEauthorrefmark{5}} 
      \IEEEauthorblockA{\IEEEauthorrefmark{1}\IEEEauthorrefmark{2}Department of EE, Indian Institute of Technology Jammu, India}
      \IEEEauthorblockA{\IEEEauthorrefmark{3}\IEEEauthorrefmark{4}\IEEEauthorrefmark{5}Wireless Communications Laboratory, Tyndall National Institute, Dublin, Ireland}

      \IEEEauthorblockA{\IEEEauthorrefmark{3}School of Computer Science and Information Technology, University College Cork, Cork, Ireland}

      \IEEEauthorblockA{\IEEEauthorrefmark{3}Trinity College Dublin, Dublin, Ireland}
  %    \IEEEauthorblockA{\IEEEauthorrefmark{2}\IEEEauthorrefmark{5}School of Electrical and Electronic Engineering, University College Dublin, Ireland}
         
  % \IEEEauthorblockA{\IEEEauthorrefmark{3}Engineering and Applied Science, Memorial University, Canada}
           
    \textrm{\{\IEEEauthorrefmark{1}shalini.tripathi,\IEEEauthorrefmark{2}ankur.bansal\}@iitjammu.ac.in},
    \textrm{\{\IEEEauthorrefmark{3}holger.claussen,\IEEEauthorrefmark{4}lester.ho,\IEEEauthorrefmark{5}chinmoy.kundu\}@tyndall.ie}
}

\maketitle
 \thispagestyle{empty}
\pagestyle{plain} 

\begin{abstract}
In this paper, a joint optimal allocation of transmit power at the source and jamming power at the destination is proposed to maximize the average secrecy energy efficiency (SEE) of a wireless network within a finite time duration. The destination transmits the jamming signal to improve secrecy by utilizing full-duplex capability. The source and destination both have energy harvesting (EH) capability with limited battery capacity. Due to the Markov nature of the system, the problem is formulated as a finite-horizon reinforcement learning (RL) problem. We propose the finite-horizon joint power allocation (FHJPA) algorithm for the finite-horizon RL problem and compare it with a low-complexity greedy algorithm (GA). An infinite-horizon joint power allocation (IHJPA) algorithm is also proposed for the corresponding infinite-horizon problem. A comparative analysis of these algorithms is carried out in terms of SEE, expected total transmitted secure bits, and computational complexity. The results show that the FHJPA algorithm outperforms the GA and IHJPA algorithms due to its appropriate modelling in finite horizon transmission. When the source node battery has sufficient energy, the GA can yield performance close to the FHJPA algorithm despite its low-complexity. When the transmission time horizon increases, the accuracy of the infinite-horizon model improves, resulting in a reduced performance gap between FHJPA and IHJPA algorithms. The computational time comparison shows that the FHJPA algorithm takes $16.6$ percent less time than the IHJPA algorithm.

\end{abstract}
\begin{IEEEkeywords}
Energy harvesting, Markov decision process, reinforcement learning, secrecy energy efficiency.
\end{IEEEkeywords}
\IEEEpeerreviewmaketitle
% \vspace{-.2cm}

%%%%%%%%%%%%%%%%%%%%%%%%%%%%%%%%%%%%%%%%%%%%%%%%%%%%%%%%%%%%%%%%%%%%%%%%%%%%%%%%%%%%%%%%%%%%%%%%%%%%

\section{Introduction}

Wireless communication networks that use energy harvesting (EH) technologies have received a lot of attention because of their ability to extend their operating lifetime. 
% Since standard battery-powered transceivers have limited energy storage, EH allows nodes to accumulate energy from ambient sources like solar, wind, vibrations, etc. 
Wireless networks with EH devices can be beneficial in various real-time applications, such as 
environmental monitoring, advanced healthcare delivery, industrial process control, vehicle tracking, smart homes, border surveillance,  etc. \cite{sah2022_WSN_intro}.  
% Unlike battery-powered devices, an EH transmitter or a receiver can theoretically continue operating over an infinite time duration. 
% However, in practice, 
% Along with the rate of EH, the channel conditions, energy available in the battery, and physical failure, also affect 
To extend the lifetime of the EH Wireless networks, the network transmission power strategy should be optimized to adapt to the random EH process, channel variations in time, and the available battery energy.

To optimize the cumulative performance of a network where optimization cannot be considered independently for each time slot (TS) due to the causality constraints on different parameters, sequential decision-making problems are formulated. The decisions are made using only past and current knowledge of the underlying stochastic processes without the knowledge of future information.
These \textit{online optimization} (stochastic) problems require full statistical knowledge along with the causal information on the realizations of the underlying stochastic processes \cite{rui2012finite}. 
Solutions to these problems can be obtained optimally through the use of the reinforcement learning (RL) technique, which uses the Markov Decision Process (MDP) framework. 
% In RL, an agent learns decision-making skills by interacting with its surroundings and getting feedback in the form of rewards \cite{sutton1998RL}. 

Based on the time horizon during which power allocation strategies are optimized for a network, the RL problems fall into two categories: finite-horizon and infinite-horizon \cite{wong2012ICC,rui2012finite,jing2017infinite, Octavia2019, kashef2012optimal,dohler2013learning,wong2014joint, shalini_TVT_arxiv }. In finite-horizon problems, the policies execute actions at a predetermined number of finite decision epochs. For example, the transmit power of a network is optimized for a finite number of TSs  \cite{wong2012ICC,rui2012finite,jing2017infinite, Octavia2019}. 
Finite-horizon modelling is most appropriate for problems where the problem setting is inherently non-stationary and varies over different time stages \cite{finitevsinfinite_Qlearning}. 
% For example, smart grid problems are inherently finite-horizon in nature as energy generation depends on the time of day or season.  
In contrast, infinite-horizon problems never terminate and provide stationary policies for the steady state. Infinite-horizon modelling is implemented in practice when the number of decision epochs is infinite or initially unknown \cite{finitevsinfinite_cpomdp}. 
Infinite-horizon modelling is used for power allocation in wireless networks, assuming the network's lifetime is infinite or stochastic, in \cite{kashef2012optimal, dohler2013learning, wong2014joint, shalini_TVT_arxiv}.

Recently, there has been an increasing focus on improving the security of the EH wireless system due to the open nature of the wireless medium \cite{SEE_RL_WCNC23, SEE_RL_VTC23, shalini_TVT_arxiv}. The wireless systems must not only be secure, it should also be energy efficient (EE) to prolong the lifetime of wireless networks \cite{octavia2020VTC}. An Unmanned aerial vehicle (UAV)-reconfigurable intelligent surface (RIS)-assisted maritime communication system under jamming attacks is considered in \cite{yang2024EH_DRL}. A deep RL (DRL)-based approach is proposed to jointly optimize the transmission power of a base station, the placement of the UAV-RIS, and the RIS's reflecting beamforming by maximizing the system EE. In \cite{SEE_RL_WCNC23}, RIS-assisted mmWave UAV communication system is considered in the presence of eavesdroppers. The worst-case secrecy energy efficiency (SEE) of the system is maximized by jointly optimizing the flight trajectory using active and passive beamforming based on RIS. In \cite{SEE_RL_VTC23}, a multi-RIS-aided simultaneous wireless information and power transfer (SWIPT) wireless network is considered. The SEE of the system is maximized by jointly optimizing the beamforming of an access point, the RIS phase shifts, and the RIS ON/OFF control.

% A DRL-based approach is proposed to jointly optimize UAV placement and
% RIS surface elements 
% under imperfect CSI conditions. 
% To tackle the problem, the authors propose a novel Twin-Twin-Delayed Deep Deterministic Policy Gradient (TTD3) algorithm.
% under imperfect CSI conditions 
% , where a multi-antenna access point serves multiple information decoding and EH receivers through direct and RIS-assisted links. 
% , while ensuring secrecy, data rate, and energy harvesting constraints. 
% The authors propose a Deep Deterministic Policy Gradient (DDPG) algorithm.

% None of the aforementioned works that address the EE of networks using RL approaches in \cite{octavia2020VTC, 
% yang2024EH_DRL, SEE_RL_WCNC23, SEE_RL_VTC23 } consider the EH capability at the source or the destination nodes, which is essential for extending the lifespan of a network.

While all the aforementioned works \cite{octavia2020VTC, yang2024EH_DRL, SEE_RL_WCNC23, SEE_RL_VTC23 } address EE using RL approaches, \cite{octavia2020VTC, yang2024EH_DRL} do not consider SEE. Though \cite{SEE_RL_WCNC23, SEE_RL_VTC23} consider SEE, \cite{SEE_RL_WCNC23} does not incorporate EH capabilities and  \cite{SEE_RL_VTC23} assumes EH only at the destination node. There is a practical need for EH at both the source and the destination nodes to enhance the network's lifetime. Recently in \cite{shalini_TVT_arxiv}, joint optimization of transmit and jamming power is carried out
using RL assuming a full-duplex destination node with EH capability at both the source and destination nodes, however, its objective is to maximize long-term expected total transmitted secure bits over the lifetime of the network rather than SEE. To the best of our knowledge, no existing work in the literature simultaneously incorporates EH at both the source and destination nodes while also maximizing SEE using RL approaches.

Motivated by the above discussion, this paper presents a novel joint power allocation strategy, optimizing both the transmit power at the source and the jamming power at the destination by leveraging the destination’s full-duplex capability to maximize SEE.  While prior studies adopt either finite \cite{wong2012ICC,rui2012finite,jing2017infinite, Octavia2019} or infinite-horizon \cite{kashef2012optimal, dohler2013learning, wong2014joint, shalini_TVT_arxiv} RL models, our work is the first to investigate and compare both in maximizing SEE in joint transmit and destination jamming power allocation problems, providing valuable insights into how optimal policies evolve with respect to varying time horizons in the considered system. Our contributions can be summarized below.
\begin{itemize}
    \item We study the optimal joint power allocation for source and destination to maximize average SEE over fixed TSs where the full-duplex destination uses jamming for secrecy. Both source and destination have EH capabilities. The problem is formulated as a finite-horizon MDP and a solution is provided using RL approaches.

    % \item We compared the finite-horizon MDP problem with the corresponding infinite-horizon MDP problem using

    \item We propose a finite-horizon joint power allocation (FHJPA) algorithm for the finite-horizon formulation and compare its performance with that of the infinite-horizon joint power allocation (IHJPA) algorithm proposed for the infinite-horizon formulation. A comparative analysis of these approaches is carried out in terms of SEE, expected total transmitted secure bits, and computation complexity by including a greedy algorithm (GA) as a baseline.

\end{itemize}

% The rest of the paper is organized as follows: Section II describes the system model. Section III formulates the problem of joint transmit and jamming power allocation for the source and destination nodes, respectively. Section IV proposes solution approaches, and Section V provides numerical results. Finally, Section VI concludes the paper. 

\textit{Notation:}  
% $||\mathbf{h}||$ denotes the Euclidean norm of a vector $\mathbf{h}$,
% % $\mathbb{N}$ is a set of natural numbers,
$\mathbb{P[\cdot]}$ denotes the probability of an event, $\mathbb{E}\left[\cdot \right]$ denotes the expectation operator.
$\max\{\cdot\}$ and $\min\{\cdot\}$ denote the maximum and minimum of its arguments, respectively.
$\mathcal{O}(\cdot)$ shows the worst case complexity of an algorithm.
%%%%%%%%%%%%%%%%%%%%%%%%%%%%%%%%%%%%%%%%%%%%%%%%%%%%%%%%%%%%%%%%%%%%%%%%%%%%%%%%%%%%%%%%%%%%%

\section{System Model}

 \begin{figure}
 \raggedleft
  \includegraphics[width=3in]{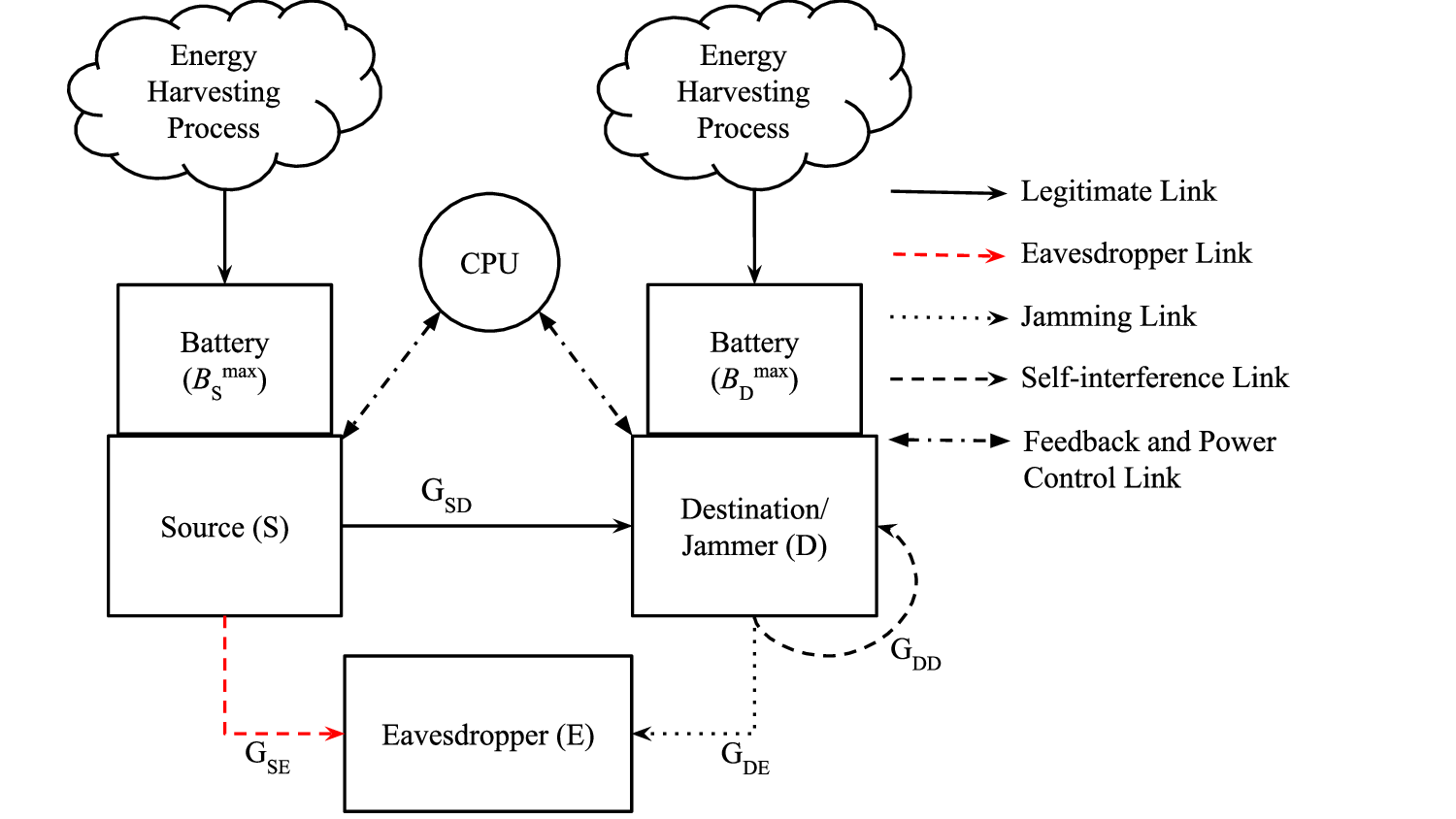}
 \caption{A wireless system with an EH source, an EH full-duplex destination, and a passive eavesdropper.}
 \label{fig_system}
 \vspace{-0.4cm}
\end{figure}
% \vspace{-1cm}

We consider a wireless communication system comprising a source node ($\textrm{S}$), a destination node ($\textrm{D}$), and a passive eavesdropper ($\textrm{E}$), as shown in Fig. \ref{fig_system}. 
% The nodes $\textrm{S}$ and $\textrm{D}$ are assumed to have EH capability. 
$\textrm{S}$ and $\textrm{D}$ are equipped with EH devices powered by rechargeable batteries with limited energy storage, while $\textrm{E}$ relies on a traditional power source. 
The communication occurs in discrete TSs, indexed by $k \in \mathcal{K} = \{0,1, \ldots, K-1\}$, where $K$ is the total number of available TSs known \textit{a priory} and each TS has a fixed duration 
of $T_s$ seconds \cite{wong2012ICC, wong2014joint, dohler2013learning, Octavia2019}. Node $\textrm{S}$ is assumed to have an uninterrupted data supply for transmission in all TSs. 
% Meanwhile, node $\textrm{D}$ operates in full-duplex mode, allowing it to receive signals from node $\textrm{S}$ while simultaneously transmitting jamming signals toward node $\textrm{E}$. However, this jamming introduces self-interference, which is mitigated through a self-interference cancellation (SIC) mechanism, though some residual interference persists. 
At the same time, $\textrm{D}$ operates in full-duplex mode. This allows $\textrm{D}$ to receive signals from $\textrm{S}$ while simultaneously transmitting jamming signals to degrade the received signals at node $\textrm{E}$. The jamming process creates self-interference at $\textrm{D}$. We assume that a self-interference cancellation (SIC)  mechanism exists at $\textrm{D}$, which reduces most of this self-interference, though some residual self-interference persists.
The effectiveness of SIC is characterized by a parameter $0\leq \alpha \leq 1$, where $\alpha = 1$ represents no SIC and $\alpha = 0$ corresponds to perfect SIC \cite{insoo2019FD}. 
Additionally, a central processing unit (CPU) is assumed to be present in the network, equipped with knowledge of global channel state information (CSI) and battery status, which it utilizes for making power allocation decisions.
% \begin{figure}
%  \raggedleft
%  \includesvg[width=3in]{system_model.svg}
%  \caption{A wireless system with an EH source, an EH full-duplex destination, and a passive eavesdropper.}
%  \label{fig_system}
%  \vspace{-0.4cm}
% \end{figure}
% % \vspace{-1cm}

% \vspace{-.4cm}

\subsection{EH Model} 
% \vspace{-.1cm}
Harvested energy at $\textrm{S}$ and $\textrm{D}$  in the $k$th TS is denoted as $H_\textrm{S}^{(k)} \in \mathcal{H}_\textrm{S}$ and $H_\textrm{D}^{(k)} \in \mathcal{H}_\textrm{D}$, respectively. Here, $\mathcal{H}_\textrm{S} = \{0,E_\textrm{S}\}$ and $\mathcal{H}_\textrm{D} = \{0,E_\textrm{D}\}$ represent the sets of feasible harvested energy levels.  
The EH process at both nodes is modelled as independent and non identically distributed Bernoulli processes with probabilities $p$ and $q$, respectively. We also assume that the EH process is independent of data transmission. Consequently, in each TS $k \in \mathcal{K}$, the probabilities of $\textrm{S}$ and $\textrm{D}$ harvesting $E_\textrm{S}$ and $E_\textrm{D}$ units of energy are given by $\mathbb{P}[H_\textrm{S}^{(k)}=E_\textrm{S}]=p$ and $\mathbb{P}[H_\textrm{D}^{(k)}=E_\textrm{D}]=q$, respectively, while no energy is harvested with probability $\mathbb{P}[H_\textrm{S}^{(k)}=0]=1-p$ and $\mathbb{P}[H_\textrm{D}^{(k)}=0]=1-q$, respectively.  

$\textrm{S}$ and $\textrm{D}$ are equipped with finite-capacity batteries, with respective capacities $B_{\textrm{S}}^{\max}$ and $B_{\textrm{D}}^{\max}$ energy units. The battery energy levels at $\textrm{S}$ and $\textrm{D}$ in the $k$th TS are denoted by $B_\textrm{S}^{(k)} \in \mathcal{B}_\textrm{S}$ and $B_\textrm{D}^{(k)} \in \mathcal{B}_\textrm{D}$, respectively, where $\mathcal{B}_\textrm{S} = \{0,1, \ldots, B_{\textrm{S}}^{\max}\}$ and $\mathcal{B}_\textrm{D} = \{0,1, \ldots, B_{\textrm{D}}^{\max}\}$ define the sets of possible discrete battery states.   

The energy used for data transmission and jamming in any TS is constrained by the available battery energy in the respective nodes in that TS. Additionally, the harvested energy cannot be stored beyond the battery's maximum capacity. The harvested energy $H_\textrm{S}^{(k)}$ in the $k$th TS together with the transmitted energy at the $k$th TS are utilized to update the battery state for the $(k+1)$th TS. Consequently, the battery states in the $(k+1)$th TS are updated as follows
{\small
\begin{align}
\label{battery_Bs}
   B_\textrm{S}^{(k+1)} 
   \hspace{-0.1cm}
   &= 
   \hspace{-0.1cm}
   \left\{ 
    \hspace{-0.2cm}
   \begin{array}{lcl}
    \min
    \{ B_\textrm{S}^{(k)} - P_\textrm{S}^{(k)}  T_s + E_\textrm{S},B_{\textrm{S}}^{\textrm{max}} \}
    \,\, \mbox{for}\,\,H_\textrm{S}^{(k)} \hspace{-0.1cm}=\hspace{-0.1cm} E_\textrm{S}  
    \\ B_\textrm{S}^{(k)} - P_\textrm{S}^{(k)} T_s 
   \hspace{2.9cm} \mbox{for}\,\,H_\textrm{S}^{(k)} = 0 
    \end{array}
    \right.  \\
\label{battery_Bd}
   B_\textrm{D}^{(k+1)} 
   \hspace{-0.1cm}
   &= 
   \hspace{-0.1cm}
   \left\{ 
    \hspace{-0.2cm}
   \begin{array}{lcl}
    \min
    \{ B_\textrm{D}^{(k)} - P_\textrm{D}^{(k)}  T_s + E_\textrm{D},B_{\textrm{D}}^{\textrm{max}} \}
    \,\, \mbox{for}\,\,H_\textrm{D}^{(k)} \hspace{-0.1cm}=\hspace{-0.1cm} E_\textrm{D} 
    \\ B_\textrm{D}^{(k)} - P_\textrm{D}^{(k)} T_s 
   \hspace{2.9cm} \mbox{for}\,\,H_\textrm{D}^{(k)} = 0 
    \end{array}
    \right. 
\end{align}
}
where $P_\textrm{S}^{(k)} \in \mathcal{P}$ and $P_\textrm{D}^{(k)} \in \mathcal{P}$ represent the power levels used by $\textrm{S}$ and $\textrm{D}$, respectively, for transmission and jamming in the $k$th TS. The set of $M$ possible transmit power levels is given by $\mathcal{P} = \{P_1,P_2, \ldots, P_M\}$, ensuring that power constraints $P_\textrm{S}^{(k)} T_s \leq B_\textrm{S}^{(k)}$ and $P_\textrm{D}^{(k)} T_s \leq B_\textrm{D}^{(k)}$ hold. The EH and data transmission operate with distinct sets of hardware, allowing these two critical processes to function independently and concurrently. 
% It is important to note that EH and data transmission occur simultaneously, and 
The EH in the $k$th TS will be available for use in the $(k+1)$th TS onward.

\subsection{Channel and Signal Transmission Model}
The channel power gain for a given link XY during the $k$th TS is represented as $G_{\textrm{XY}}^{(k)}$, where $\textrm{XY} \in\{\textrm{SD, SE, DD, DE}\}$ corresponds to the links between nodes $\textrm{X}\in\{\textrm{S, D}\}$ and $\textrm{Y}\in\{\textrm{D, E}\}$. The self-interference link at $\textrm{D}$ is denoted by $\textrm{DD}$. The values of $G_{\textrm{XY}}^{(k)}$ are quantized into $L$ discrete levels such that $G_{\textrm{XY}}^{(k)} \in \mathcal{G}$, where $\mathcal{G} = \{G_1, G_2, \ldots, G_L\}$. Within a given TS, the channel power gain remains constant but transitions to a new value in the next TS according to the predefined set $\mathcal{G}$. This transition is assumed to follow a first-order Markov process \cite{dohler2013learning}, which models the uncertainty of wireless communication channels.

The received signals at $\textrm{D}$ and $\textrm{E}$ during the $k$th TS can be expressed as
{\small
\begin{align}
\label{EQ_receved_signal}
    y_\textrm{D}^{(k)} &= \sqrt{G_{\textrm{SD}}^{(k)} P_\textrm{S}^{(k)}} x_\textrm{S}^{(k)} +  \sqrt{\alpha G_{\textrm{DD}}^{(k)} P_\textrm{D}^{(k)}} w_\textrm{D}^{(k)} + z_\textrm{D}^{(k)},\\    
    y_\textrm{E}^{(k)} &= \sqrt{G_{\textrm{SE}}^{(k)} P_\textrm{S}^{(k)}} x_\textrm{S}^{(k)} + \sqrt{G_{\textrm{DE}}^{(k)} P_\textrm{D}^{(k)}} w_\textrm{D}^{(k)} + z_\textrm{E}^{(k)},
\end{align}
}
where $x_\textrm{S}^{(k)}$ and $w_\textrm{D}^{(k)}$ denote the unit energy information and jamming signals transmitted by $\textrm{S}$ and $\textrm{D}$, respectively. The term $\alpha$ accounts for self-interference attenuation, and $z_\textrm{D}^{(k)}$, $z_\textrm{E}^{(k)}$ are additive white Gaussian noise (AWGN) components at $\textrm{D}$ and $\textrm{E}$ respectively, with mean zero and power spectral density $N_0$ W/Hz.
The corresponding signal-to-interference-plus-noise ratios (SINRs) at $\textrm{D}$ and $\textrm{E}$ in the $k$th TS are given by 
{\small
\begin{align}
\gamma_\textrm{D}^{(k)} =\frac{G_{\textrm{SD}}^{(k)} P_\textrm{S}^{(k)}}{\alpha P_\textrm{D}^{(k)} G_{\textrm{DD}}^{(k)} + W  N_0},
% \end{align}
% and
% \begin{align}
 ~~~\gamma_\textrm{E}^{(k)} =\frac{G_{\textrm{SE}}^{(k)} P_\textrm{S}^{(k)}}{P_\textrm{D}^{(k)} G_{\textrm{DE}}^{(k)} + W N_0},   
\end{align}
}
respectively, where $W$ is the bandwidth of the channel.

\subsection{Performance Metric}
We define  SEE as the performance metric for which we first define the achievable secrecy rate. The achievable secrecy rate of the network, measured in bits per second (bps), during the $k$th TS is determined by the difference in achievable rates between the destination and the eavesdropper. It is expressed as
{\small
\begin{align}
        % C_S = [C_D - C_E]^+
    C_\textrm{S}^{(k)} = \max \{C_\textrm{D}^{(k)} - C_\textrm{E}^{(k)},0\}\quad \text{bps},
    \label{eqCSk}
\end{align}
}
where $ C_\textrm{D}^{(k)} = W \log_2{(1+\gamma_\textrm{D}^{(k)})}$ and $C_\textrm{E}^{(k)} = W \log_2{(1+\gamma_\textrm{E}^{(k)})}$ denote the achievable rates for the destination and eavesdropping channels during the $k$th TS, respectively. The function $\max\{\cdot\}$ in (\ref{eqCSk}) ensures that the secrecy rate remains non-negative. 

The average SEE is defined as the average number of secure bits transmitted per energy unit over $K$ TSs and is expressed as \cite{octavia2020VTC} 
{\small
\begin{align}
        \eta_{E}
    %     =                \frac{C_\textrm{S}}
    % {\sum_{k=0}^{K-1}(P_\textrm{S}^{(k)}+P_\textrm{D}^{(k)}) T_x}
    =    \mathbb{E} \lb[ \frac{1}{K} \sum_{k=0}^{K-1}
        \frac{C_\textrm{S}^{(k)} T_s}
    {(P_\textrm{S}^{(k)}+P_\textrm{D}^{(k)})T_s}\rb],
        \label{eq_SEE}
    \end{align}   
    }
    where $\mathbb{E}[\cdot]$ represents the expectation over all stochastic variables, including $G_{\textrm{XY}}^{(k)}$ for all $\textrm{XY} \in\{\textrm{SD, SE, DD, DE}\}$, as well as $H_{\textrm{S}}^{(k)}$ and $H_{\textrm{D}}^{(k)}$.  
% A higher value of $\eta_{E}$ signifies a more secrecy energy efficient secure communication system, ensuring better utilization of available energy. 
%%%%%%%%%%%%%%%%%%%%%%%%%%%%%%%%%%%%%%%%%%%%%%%%%%%%%%%%%%%%%%%%%%%%%%%%%%%%%%%%%%%%%%%%%%%%%%%%
 % \vspace{-0.75cm}
\section{Problem Formulation}
 % \vspace{-0.5cm}
The objective of the problem formulation is to maximize the average SEE over $K$ TSs in (\ref{eq_SEE}), where $K$ is known \textit{a priory}, by optimally allocating $P_\textrm{S}^{(k)}$ and $P_\textrm{D}^{(k)}$ in each TS.
% The solution should consider not only the current information, such as current battery energy level, EH rate, channel condition, and self-interference attenuation factor, but also the probability that the network remains operational at each TS.
Accordingly, we formulate the problem of finding the joint power allocation for transmitting and jamming power as
\begin{subequations}
{\small
\begin{align}
    \text{P1}:  \underset{\{P^{(k)}_\textrm{S},P^{(k)}_\textrm{D}\}_{k = 0}^K }{\text{maximize}} 
     &  \eta_{E}\\
     & \text{s.t.}\,
      (\ref{battery_Bs}), (\ref{battery_Bd})\label{joint_optimization_b}\\
     & 0  \leq P_\textrm{S}^{(k)} \leq {B_{\textrm{S}}^{(k)}}/{ T_s} \label{joint_optimization_c}\\
     & 0  \leq P_\textrm{D}^{(k)} \leq {B_{\textrm{D}}^{(k)}}/{T_s}. \label{joint_optimization_d}
\end{align}
}
\label{Problem}
\end{subequations}
The constraint associated with updating the battery states at $\textrm{S}$ and $\textrm{D}$ in each TS, considering the transmit power and harvested energy, is expressed in (\ref{joint_optimization_b}). The transmit and jamming power constraints are respectively expressed in (\ref{joint_optimization_c}) and (\ref{joint_optimization_d}). 

% \color{red}
% An insightful observation of the problem P1 reveals that the joint optimal allocation of powers at $\textrm{S}$ and $\textrm{D}$ depends not only on the knowledge of channel conditions and battery levels in the current TS, it also depends on their values in the future.
As our system follows the Markov property, the formulated problem in (\ref{Problem}) is an online sequential decision-making problem with finite action and state spaces with a bounded and consistent immediate reward function. Therefore, we use the RL framework, which is essentially an MDP, to obtain the optimal policy in each decision epoch to maximize the average reward over finite $K$ number of TSs \cite{puterman2014markov}.

% \color{blue}
% It is interesting to note in (\ref{joint_optimization_a}) that if no energy is used over the lifetime $K$, no secure bits are transmitted. This ensures that the metric does not allow a strategy to save energy by not using any at all, as that would not maximize (\ref{joint_optimization_a}).
% \color{black}

We require an MDP model for the RL-based solution which has a tuple consisting of states, actions, rewards, and transition probabilities.
Assume that the state in $k$th TS 
is defined as $s^{(k)}=(G_{\textrm{SD}}^{(k)}, G_{\textrm{SE}}^{(k)}, G_{\textrm{DD}}^{(k)}, G_{\textrm{DE}}^{(k)}, \allowbreak B_\textrm{S}^{(k)}, B_\textrm{D}^{(k)})$. The state space is given by  $\mathcal{S} = \mathcal{G}_{\textrm{SD}} \times \mathcal{G}_{\textrm{SE}} \times \mathcal{G}_{\textrm{DD}} \times \mathcal{G}_{\textrm{DE}} \times \mathcal{B}_\textrm{S} \times \mathcal{B}_\textrm{D}$ with finite number of discrete possible states $N_S$ (the cardinality of $\mathcal{S}$). An action $a^{(k)}$ is taken in the $k\in\{1,2,\ldots, K\}$th TS to optimize the problem P1, i.e., a pair of transmit powers $\{P_\textrm{S}^{(k)},P_\textrm{D}^{(k)}\}$ is decided in each TS from the feasible action set $U(s^{(k)})$ such that $a^{(k)} \in U(s^{(k)})$ where
% \begin{align}
% \label{eq_feasible_action}
% U(s^{(k)}) = \lb\{P_S^{(k)}, P_D^{(k)} \,\middle|\, 0\leq P_S^{(k)}\leq \frac{B_S^{(k)}}{T_s},\ 0\leq P_D^{(k)}\leq \frac{B_D^{(k)}}{T_s} \rb\}
% \end{align}
{\small
\begin{align}
\label{eq_feasible_action}
      U(s^{(k)}) = \bigg\{P_\textrm{S}^{(k)},P_\textrm{D}^{(k)} \mid 0 \leq P_\textrm{S}^{(k)} \leq \frac{B_{\textrm{S}}^{(k)}}{T_s}, 0 \leq P_\textrm{D}^{(k)} \leq \frac{B_{\textrm{D}}^{(k)}}{T_s}\bigg\}. 
\end{align}
}
The action $a^{(k)}$ belongs to the set $\mathcal{A} =\{\delta_1,\ldots, \delta_{ N_A }\}$ of all possible actions where an action $\delta_i$ for any $i\in\{1,\ldots, N_A\}$ is the pair of transmit power levels $\{ P _m \in \mathcal{P},  P _n\in \mathcal{P}\}$ for any $m,n\in\{1,\ldots, M\}$, and $N_A =  M^2$ is the total number of possible actions. State transition probability of transitioning to the state $s^{(k+1)}$ in the $(k+1)$th TS from the state $s^{(k)}$ by taking an action $a^{(k)}$ in the $k$th TS is expressed as 
{\small
\begin{align}
    &\mathbb{P}[s^{(k+1)}\mid s^{(k)}, a^{(k)}]
    % \nn\\ 
 %    & = \mathbb{P}[G_{\textrm{SD}}^{(k+1)}, G_{\textrm{SE}}^{(k+1)}, G_{\textrm{DD}}^{(k+1)}, G_{\textrm{DE}}^{(k+1)}, 
 % B_\textrm{S}^{(k+1)}, B_\textrm{D}^{(k+1)}\mid \nn\\ 
 %    &~~~~G_{\textrm{SD}}^{(k)}, G_{\textrm{SE}}^{(k)}, G_{\textrm{DD}}^{(k)}, G_{\textrm{DE}}^{(k)}, B_\textrm{S}^{(k)}, B_\textrm{D}^{(k)}, P_\textrm{S}^{(k)}, P_\textrm{D}^{(k)}] \nn\\ 
    % &
    = \mathbb{P}[G_{\textrm{SD}}^{(k+1)}\mid G_{\textrm{SD}}^{(k)}] \times \mathbb{P}[G_{\textrm{SE}}^{(k+1)}\mid G_{\textrm{SE}}^{(k)}] \nn \\
    &~~~~\times \mathbb{P}[G_{\textrm{DD}}^{(k+1)}\mid G_{\textrm{DD}}^{(k)}] 
    \times \mathbb{P}[G_{\textrm{DE}}^{(k+1)}\mid G_{\textrm{DE}}^{(k)}] \nn \\
    &~~~~\times \mathbb{P}[B_\textrm{S}^{(k+1)} \mid B_\textrm{S}^{(k)}, H_\textrm{S}^{(k)}, P_\textrm{S}^{(k)}]
    \times\mathbb{P}[H_\textrm{S}^{(k)}]  \nn\\
    &~~~~\times \mathbb{P}[B_\textrm{D}^{(k+1)} \mid B_\textrm{D}^{(k)},  H_\textrm{D}^{(k)} , P_\textrm{D}^{(k)}]\times\mathbb{P}[H_\textrm{D}^{(k)}],
    \label{Transition_Prob}
\end{align} 
}
where $\mathbb{P}[B_\textrm{S}^{(k+1)} \mid B_\textrm{S}^{(k)}, H_\textrm{S}^{(k)}, P_\textrm{S}^{(k)}]$ and $\mathbb{P}[B_\textrm{D}^{(k+1)} \mid B_\textrm{D}^{(k)}, H_\textrm{D}^{(k)}, P_\textrm{D}^{(k)}]$ are equal to 1 if (\ref{battery_Bs}) and (\ref{battery_Bd}) are satisfied, zero otherwise. We also have $\mathbb{P}[H_\textrm{S}^{(k)}]=p$ when  $H_\textrm{S}^{(k)}=E_S$, $\mathbb{P}[H_\textrm{S}^{(k)}]=1-p$ when  $H_\textrm{S}^{(k)}=0$, $\mathbb{P}[H_\textrm{D}^{(k)}]=q$ when  $H_\textrm{D}^{(k)}=E_D$, and $\mathbb{P}[H_\textrm{D}^{(k)}]=1-q$ when  $H_\textrm{D}^{(k)}=0$.
% \textcolor{red}{As you are saying that if eqn 1 and 2 are satisfied, I think it is better to delete with probability p and q thing in 1 and 2. Because it may be confusing. in 1 and 2, you can just write ``when energy harvested $H_\textrm{S}^{(k)} = E_H$'' and ``when not harvested''}. 
If $\mathbb{P}[B_\textrm{S}^{(k+1)} \mid B_\textrm {S}^{(k)}, H_\textrm{S}^{(k)}, P_\textrm{S}^{(k)}]$ and $\mathbb{P}[B_\textrm{D}^{(k+1)} \mid B_\textrm{D}^{(k)}, H_\textrm{D}^{(k)}, P_\textrm{D}^{(k)}]$ are equal to zero, then (\ref{Transition_Prob}) also becomes zero indicating the impossibility of a transition from state $s^{(k)}$ to state $s^{(k+1)}$ while taking action $a^{(k)}$. An action $a^{(k)}$ also results in an immediate reward $R^{(k)}(s^{(k)},a^{(k)})$. In the context of our problem, the immediate reward function in the $k$th TS from (\ref{eqCSk}) is
{\small
\begin{align}
    R^{(k)}(s^{(k)},a^{(k)}) = \frac{   C_\textrm{S}^{(k)} T_s}
    {%\sum_{k=0}^{K-1}
    (P_\textrm{S}^{(k)}+P_\textrm{D}^{(k)})T_s}  = \frac{   C_\textrm{S}^{(k)}}
    {%\sum_{k=0}^{K-1}
    P_\textrm{S}^{(k)}+P_\textrm{D}^{(k)}}
    \label{reward}
\end{align}
}
and the average reward is expressed in (\ref{eq_SEE}). 
%%%%%%%%%%%%%%%%%%%%%%%%%%%%%%%%%%%%%%%%%%%%%%%%%%%%%%%%%%%%%%%%%%%%%%%%%%%%%%%%%%%%%%%%%%%%%%%%%%%%%%%%
\section{Proposed Solutions}
% \vspace{-.1cm}
In this section, we provide an optimal and a low-complexity sub-optimal solution for the proposed problem in (\ref{Problem}). The optimal solution to the problem is provided with the help of the backward induction (BI) algorithm \cite{puterman2014markov}, denoted as the finite-horizon joint power allocation (FHJPA) algorithm in this paper. The sub-optimal solution is a greedy one, as explained later in this section. 
These finite-horizon solutions are compared with an infinite-horizon solution, denoted as the infinite-horizon joint power allocation (IHJPA) algorithm. All the solution methodologies are described in detail in the following subsections.

\subsection{Finite-Horizon Joint Power Allocation (FHJPA)} 
The FHJPA algorithm is described in Algorithm \ref{Algo_FHJPA}. For its execution, the algorithm requires the value function for each state-action pair $(s^{(k)},a^{(k)})$ for the $k$th TS for any $s^{(k)} \in \mathcal{S}$ and $a^{(k)}\in\mathcal{A}$. The value function is the sum of the immediate reward and the expected future reward when the action $a^{(k)}$ is taken in the state $s^{(k)}$ and is mathematically defined by Bellman's equation in (\ref{eq:V_K+1}) \cite{puterman2014markov}. 
% designed to determine \textcolor{red}{the optimal stationary deterministic policy $d^*(s)$ over a finite number of TSs, % which is based on backward induction (BI) \cite{Octavia2019}. 
The FHJPA algorithm essentially utilises the well-known BI algorithm, which starts iterating from the last TS, as shown in line 1, and moves sequentially toward the first TS. In the $k$th iteration, the algorithm finds the maximum value function $V(s^{(k)})$ for each $s^{(k)} \in \mathcal{S}$ as in (\ref{eq:V_K+1}) and its corresponding optimal policy $d^*(s^{(k)})$, i.e. a pair of optimal transmit powers, as in (\ref{eq:policyK+1}) \cite{puterman2014markov}. 
The maximum value function in (\ref{eq:V_K}) is simply the maximum immediate reward as no future reward is possible for the last TS. 
The optimal policy $d^*(s^{(k)})$ corresponding to each $s^{(k)} \in \mathcal{S}$ is stored in a look-up table. 

In the \textit{Transmission Phase}, as described in Algorithm \ref{transmission_phase}, the power allocation for each TS is implemented by directly retrieving the actions corresponding to the state in that TS from the look-up table.

\begin{algorithm} 
\scriptsize   
% \footnotesize
% \begin{tiny}
\caption{\textbf{Algorithm 1: FHJPA}}
\label{Algo_FHJPA}
    \hspace*{\algorithmicindent} \textbf{Input:} Set of states, actions, state transition probabilities, and rewards. \\
    \hspace*{\algorithmicindent} \textbf{Output:} Optimal policy $d^*(s)$.
    \begin{algorithmic}[1]
    \State Set $k = K$ and compute $V(s^{(k)})$ and $d^{*}(s^{(k)})$, for all $s^{(k)} \in \mathcal{S}$ by
         \begin{align}
            V(s^{(k)}) = \max_{a^{(k)} \in U(s^{(k)})} R(s^{(k)},a^{(k)}),  \label{eq:V_K} \\
     d^*(s^{(k)}) = \arg\max_{a^{(k)} \in U(s^{(k)})} R(s^{(k)},a^{(k)}). \label{eq:policyK}
        \end{align}
        \State Set $k = k - 1$ and compute $V(s^{(k)})$ and $d^{*}(s^{(k)})$, for all $s^{(k)} \in \mathcal{S}$ by
        \begin{align}
            V(s^{(k)}) &= \max_{a^{(k)} \in U(s^{(k)})}\Big[ R(s^{(k)},a^{(k)}) + \sum_{s'\in\mathcal{S}} \mathbb{P}(s'|s^{(k)},a^{(k)})\,V(s') \Big], \label{eq:V_K+1} \\
            d^*(s^{(k)}) &= \mathop{\mathrm{argmax}}\limits_{a^{(k)} \in U(s^{(k)})}\Big[ R(s^{(k)},a^{(k)}) + \sum_{s'\in\mathcal{S}} \mathbb{P}(s'|s^{(k)},a^{(k)})\,V(s') \Big]. \label{eq:policyK+1}
        \end{align}
        \State If $k = 1$, stop. Otherwise, return to step 4.
\end{algorithmic}
% \end{tiny}
 % \vspace{-0.25cm}
\end{algorithm}
\begin{algorithm} 
\scriptsize
% \footnotesize
\label{transmission_phase}
\caption{\textbf{Algorithm 2: Transmission Phase}}
\hspace*{\algorithmicindent} \textbf{Input}: Optimal policy $d^*(s)$.\\
\hspace*{\algorithmicindent} \textbf{Output}: Average SEE over $K$ TSs. 
\begin{algorithmic}[1]
% \State \textbf{Transmission Phase:} 
    \State Set $\eta_{E} = 0$
    \State Set $k = 0$
    \While{$k \leq K-1$}
    \State Track channel states $G_{\textrm{SD}}^{(k)}, G_{\textrm{SE}}^{(k)}, G_{\textrm{DD}}^{(k)}$ and $G_{\textrm{DE}}^{(k)} $ 
    \State Track available battery $B_\textrm{S}^{(k)}$ and $B_\textrm{D}^{(k)}$ 
    \State Set $s^{(k)} = (G_{\textrm{SD}}^{(k)}, G_{\textrm{SE}}^{(k)}, G_{\textrm{DD}}^{(k)}, G_{\textrm{DE}}^{(k)}, B_\textrm{S}^{(k)}, B_\textrm{D}^{(k)})$
    \State Obtain $a^{(k)} = (P_\textrm{S}^{(k)}, P_\textrm{D}^{(k)})$ from look-up table for state $s^{(k)}$
    \State Consume $P_\textrm{S}^{(k)}$ and $P_\textrm{D}^{(k)}$ for transmission and jamming respectively.
    \State Calculate the average SEE as in (\ref{reward})  for state $s^{(k)}$.
    \State Update battery $B_\textrm{S}^{(k)}$ and $B_\textrm{D}^{(k)}$ using (\ref{battery_Bs}) and (\ref{battery_Bd}) respectively
    \State $\eta_{E}  = \eta_{E}  + \frac{   C_\textrm{S}^{(k)}}
    {P_\textrm{S}^{(k)}+P_\textrm{D}^{(k)}}$
    \State Set $k = k + 1$
    \EndWhile
    \State $\eta_{E}  = \frac{\eta_{E}}{K} $    
\end{algorithmic}  
\label{transmission_phase}
\end{algorithm}

 % \vspace{-0.5cm}
\subsection {Greedy Algorithm (GA)}
In each TS, the GA chooses the action that maximizes the immediate reward. Thus, the power is allocated as 
\begin{align}
     a^{(k)} = \mathop{\mathrm{argmax}}\limits_{a^{(k)} \in U(s^{(k)})} R^{(k)}(s^{(k)}, a^{(k)}).  
     \label{GA_algorithm}
\end{align}

\subsection{Infinite-Horizon Joint Power Allocation (IHJPA)}
\label{IHJPA}
% $\Gamma \in [0,1)$
The IHJPA algorithm, which is essentially an infinite-horizon RL algorithm, can be implemented to maximize the average SEE in (\ref{Problem}) when $K$ is a random variable, not known \textit{a priory}. 
In an infinite-horizon problem, $K$ depends on the discount factor $\Gamma$
which is the probability that the network remains operational in a TS \cite{puterman2014markov,shalini_TVT_arxiv}.
% with mean $1/(1-\Gamma)$ can be implemented by the IHJPA as proposed in. 
% $\Gamma$ is known as the discount factor in infinite-horizon RL \cite{puterman2014markov},  
The well-known policy iteration (PI) algorithm can be utilized to develop the IHJPA algorithm. 
The PI algorithm implements \textit{Bellman's equation of optimality} \cite[pp.~174]{puterman2014markov}. We have implemented Algorithm 1: \textit{Planning Phase} of \cite{shalini_TVT_arxiv} for the IHJPA algorithm. 
% For more details on the implementation of the PI algorithm and the \textit{Planning Phase}, the readers are referred to Algorithm 1 of \cite{shalini_TVT_arxiv}. 
The IHJPA algorithm provides the optimal policy in a look-up table. Then the \textit{Transmission Phase} needs to be implemented for the power allocations in each TS.

%%%%%%%%%%%%%%%%%%%%%%%%%%%%%%%%%%%%%%%%%%%%%%%%%%%%%%%%%%%%%%%%%%%%%%%%%%%%%%%%%%%%%%%%%%%%%%%%%%
\section{Results and Discussions}
% \vspace{-.1cm}
In this section, we compare the performance of FHJPA, GA, and IHJPA algorithms in terms of the average SEE and expected total transmitted secure bits. The expected total transmitted secure bits for $K$ TSs is defined as 
% \begin{align}
    $\mu =  \mathbb{E}    \left[ \sum_{k=0}^{K-1}   C_\textrm{S}^{(k)} T_s \right] \text{bits}$, 
    % \label{eqECS}
% \end{align} 
where the expectation is taken over all the stochastic variables, including the expectation over the random variable $K$ in the case of the IHJPA algorithm. For a fair comparison between FHJPA and IHJPA algorithms, the mean value of the total number of available TSs in IHJPA algorithm is assumed to be the same as the total number of TSs available in FHJPA algorithm. In this case $K = 1/(1-\Gamma)$, where $1/(1-\Gamma)$ is the mean of the total number of available TSs in the IHJPA algorithm. 
% \textcolor{blue}{Need to write more on simulation setup, how IHJPA is simulated and compared with FHJPA.}

Unless otherwise specified, the system parameters considered for the simulation are $W = 2$ MHz, $N_0 = 10^{-20.4}$ W/Hz, $\alpha = 10^{-5}$, $T_S = 5$ ms, $E_\textrm{S} = 2$ and $E_\textrm{D} = 2$ energy units, where one energy unit corresponds to $2.5$ $\mu$J, the probabilities of harvesting $E_S = 2$ and $E_D = 2$ energy units are $p = 0.5$ and $q=0.5$, respectively,
$B_{\textrm{S}}^{\textrm{max}}$ = $B_{\textrm{S}}^{\textrm{max}}$= 5 energy units, $\mathcal{P}=\{0, 0.5, 1, 2\}$ mW,
% which corresponds to transmit and jamming energy levels of $\mathcal{E_U}=\{0,1,2,4\}$ energy units. 
the channel states are assumed to be $\{G_1, G_2\} = \{1.655 \times 10^{-13}, 3.311 \times 10^{-13}\}$ where the channel state transition probabilities are $\mathbb{P}(G_{\textrm{1}}\mid G_{\textrm{1}}) = \mathbb{P}(G_{\textrm{2}}\mid G_{\textrm{2}}) = 0.9$, and $ \mathbb{P}(G_{\textrm{2}}\mid G_{\textrm{1}}) = \mathbb{P}(G_{\textrm{1}}\mid G_{\textrm{2}}) = 0.1$.
The initial state $s^{(0)}$ is defined as $(G_2,G_2,G_2,G_2, B_{\textrm{S}}^{\textrm{max}},B_{\textrm{D}}^{\textrm{max}})$. The parameter values are considered from \cite{dohler2013learning}. 
\begin{figure}
  \centering
  \begin{subfigure}[b]{0.8\columnwidth}
    \centering
    \includegraphics[width=\textwidth]{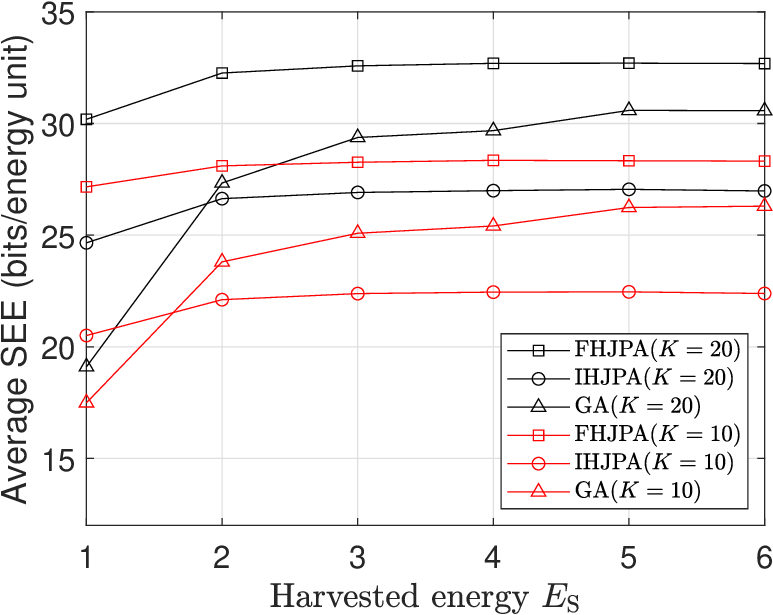}
    \caption{}
    % \vspace{-0.1cm}
    \label{fig_AvgEE_vs_EH_S_allK}
  \end{subfigure}
  \hfill
  \begin{subfigure}[b]{0.8\columnwidth}
    \centering
    \includegraphics[width=\textwidth]{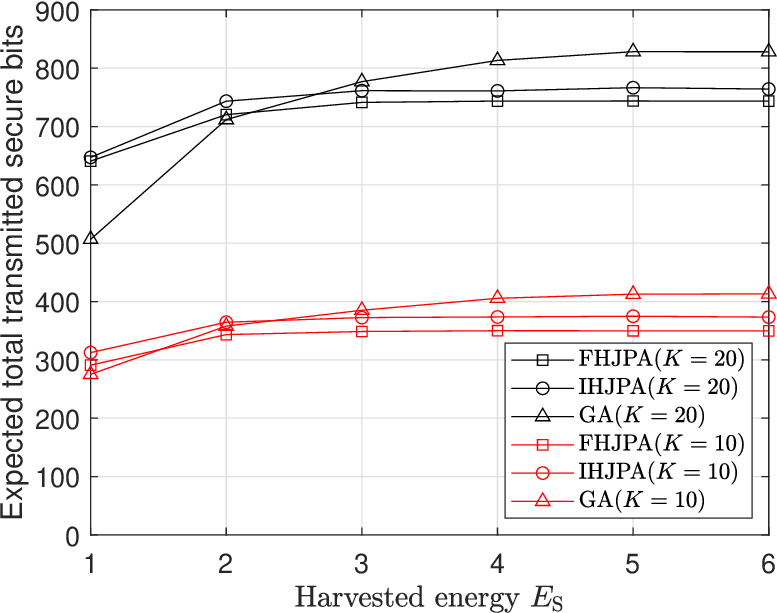}
    \caption{}
    % \vspace{-0.1cm}
    \label{fig_ETD_vs_EH_S}
  \end{subfigure}
  \caption{Average SEE and expected total transmitted secure bits versus $E_\textrm{S}$ for $K = 10$ and $K = 20$.}
  \label{fig_AvgEE_ETD_vs_EH_S}
  % \vspace{-0.5cm}
\end{figure}
% \vspace{-1cm}

In Figs. \ref{fig_AvgEE_vs_EH_S_allK} and \ref{fig_ETD_vs_EH_S}, we compare the average SEE and expected total transmitted secure bits versus $E_S$ for different values of $K$, respectively. It is observed in Fig. \ref{fig_AvgEE_vs_EH_S_allK} that the FHJPA algorithm significantly outperforms the GA and IHJPA algorithms.
As $K$ is a known fixed value, the finite-horizon MDP modelling for the proposed problem is accurate and hence, the FHJPA algorithm outperforms the IHJPA algorithm. Unlike the FHJPA algorithm, the GA only maximizes the immediate reward without considering the future reward, hence, the GA performs worse than the FHJPA algorithm. However, as $E_S$ increases, the performance of the GA gradually tends toward the performance of the FHJPA algorithm. Furthermore, the GA outperforms the IHJPA algorithm beyond a certain $E_S$.  The poor performance of the IHJPA algorithm is due to the MDP modelling of the problem through infinite-horizon RL when $K$ is low. We also find that the average SEE saturates as $E_S$ increases. The higher value of $E_S$ results in more available energy at $\textrm{S}$, which tends to fill the battery to its maximum value and results in performance saturation. The figure also depicts that as the total number of TSs increases from $K = 10$ to $K = 20$, the average SEE increases for all algorithms. 

When we compare the expected total transmitted secure bits in Fig. \ref{fig_ETD_vs_EH_S}, we find that the FHJPA algorithm is the worst and the GA outperforms both FHJPA and IHJPA algorithms after a certain $E_S$. As the aim of the optimization is to maximize the average SEE, the FHJPA algorithm reduces the transmit power which results in this observation.  On the other hand, as $E_S$ becomes higher, the GA can afford to allocate more power in each TS which leads to better expected total transmitted secure bits, however, this also lowers the average SEE of the GA as observed in Fig. \ref{fig_AvgEE_vs_EH_S_allK}. 
Figs. \ref{fig_AvgEE_vs_EH_S_allK} and \ref{fig_ETD_vs_EH_S} also indicate that optimization of the average SEE may not lead to better total transmitted secure bits. Hence, the system optimization should consider a tradeoff between the average SEE and total transmitted secure bits.

\begin{figure}
  \centering  \includegraphics[width=0.8\columnwidth]{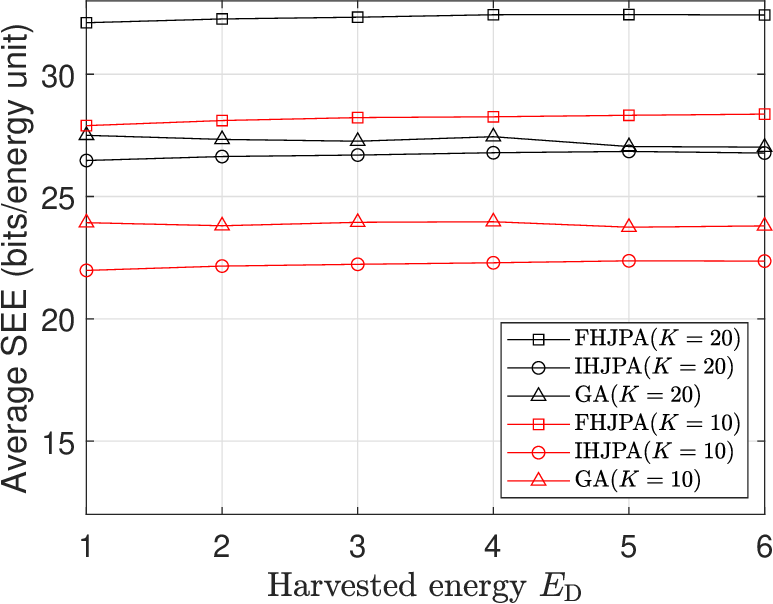}
  \caption{Average SEE versus $E_\textrm{D}$ for $K = 10$ and $K = 20$.}
  % \vspace{-0.2cm}
  \label{fig_AvgEE_vs_EH_D}
  % \vspace{-0.2cm}
\end{figure}

In Fig. \ref{fig_AvgEE_vs_EH_D}, we replicate the performances of Fig. \ref{fig_AvgEE_vs_EH_S_allK} by varying $E_D$. The observations regarding the relative performance of algorithms and the performance concerning $K$ are similar to those in Fig. \ref{fig_AvgEE_vs_EH_S_allK}.  Further, it is noticed that the increase in $E_D$ has a negligible impact on the average SEE performance of all the algorithms. 
The expected total transmitted secure bits performance by varying $E_D$ is similar to that in Fig. \ref{fig_ETD_vs_EH_S}, hence, not shown to avoid unnecessary repetition.
\begin{figure}
  \centering
  \begin{subfigure}[t]{0.8\columnwidth}
    \centering
    \includegraphics[width=\textwidth]{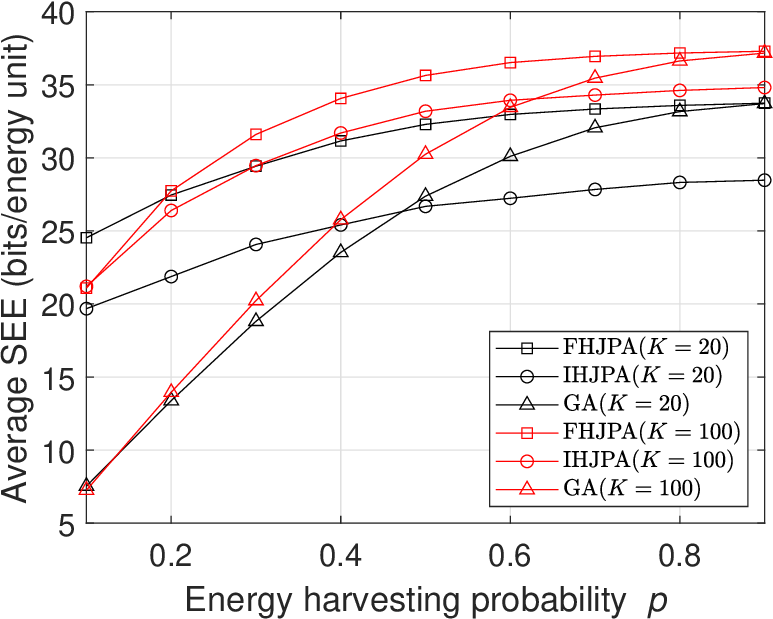}
    \caption{When $q=0.5$.}
    % \vspace{-0.1cm}
    \label{fig_AvgEE_vs_EH_varyp}
  \end{subfigure}
  \hfill
  \begin{subfigure}[t]{0.8\columnwidth}
    \centering
    \includegraphics[width=\textwidth]{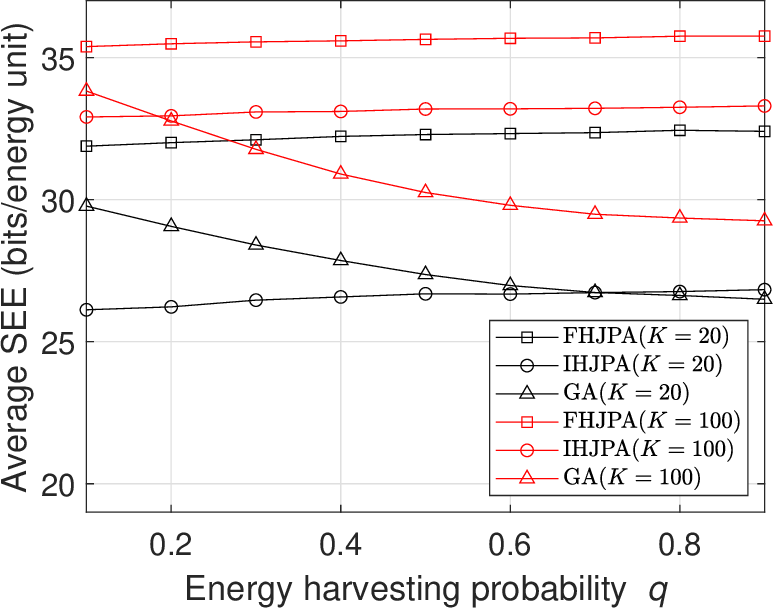}
    \caption{When $p=0.5$.}
    % \vspace{-0.1cm}
    \label{fig_AvgEE_vs_EH_varyq}
  \end{subfigure}
  \caption{Average SEE versus EH probability $p$ and $q$ where $K = 20$.}
  \label{fig_AvgEE_vs_EH_Joint_varypq}
  % \vspace{-0.5cm}
\end{figure}

In Fig. \ref{fig_AvgEE_vs_EH_varyp} and Fig. \ref{fig_AvgEE_vs_EH_varyq}, we compare the average SEE of the algorithms versus $p$ and $q$, respectively. In Fig. \ref{fig_AvgEE_vs_EH_varyp}, the average SEE gradually improves and saturates as $p$ increases. 
A higher $p$ leads to increased energy availability at $\textrm{S}$, which tends to fill the battery to its maximum capacity. The greater amount of available energy in the battery tends to  
% better average SEE for all algorithms and 
saturate the average SEE. 
Interestingly, as in Fig. \ref{fig_AvgEE_vs_EH_S_allK}, the GA performance gradually reaches the performance of the FHJPA algorithm here as well. Thus, it can be concluded that when the battery at $\textrm{S}$ has sufficient energy due to the increased amount of harvested energy or the probability of harvesting, the performance of the GA and FHJPA algorithms becomes similar. This implies that the maximization of the immediate reward in each TS becomes nearly optimal in the finite horizon when there is enough energy in the battery of $\textrm{S}$.  We also find that as $K$ increases from $20$ to $100$, the performance gap between FHJPA and IHJPA algorithms decreases. As $K$ increases, the accuracy of infinite-horizon modelling improves, hence, the performance gap decreases. 

In Fig. \ref{fig_AvgEE_vs_EH_varyq}, the increase in $q$ has a negligible effect on the average SEE performance of FHJPA and IHJPA algorithms. However, the performance of the GA degrades with increasing $q$. 

From Figs. \ref{fig_AvgEE_ETD_vs_EH_S},  \ref{fig_AvgEE_vs_EH_D}, and \ref{fig_AvgEE_vs_EH_Joint_varypq} it can be inferred that increasing the harvested energy and probability of harvesting at both $\textrm{S}$ and $\textrm{D}$ positively impact both the average SEE and expected total transmitted secure bits for all the algorithms (except for the GA in Fig. \ref{fig_AvgEE_vs_EH_varyq} only). 
{the increase in these parameters at \textrm{S} has a greater impact on improving the system performance as compared to the increase in these parameters at \textrm{D}.

% \begin{figure}
%   \centering
%   \includegraphics[width=0.6\columnwidth]{Fig_AvgEE_vs_self_interference_joint.eps}
%   \caption{Average SEE versus $\alpha$ when $K = 20$.}
%   \label{fig_AvgEE_vs_self_interference}
% \end{figure}

% In Fig. \ref{fig_AvgEE_vs_self_interference}, we compare the average SEE versus $\alpha$. The average SEE performance remains the same for most of the $\alpha$ values but as it approaches unity the performance degrades for all algorithms except in the case of the GA.   
% For sufficiently small values of $\alpha$, its effect on the average SEE is negligible, as self-interference is negligible compared to the AWGN at node $\textrm{D}$. In contrast, as $\alpha$ increases, the self-interference becomes comparable to the AWGN at node $\textrm{D}$, hence, the performance of FHJPA and IHJPA algorithms deteriorates. However, the performance of the GA improved.  This observation is due to the low power assignment for the destination jamming which in turn reduces the self-interference at the destination. 

\begin{table}
\centering
\small
\begin{tabular}{|p{1.8cm}|p{3.5cm}|p{2cm}| }
%\begin{tabular}{ |c|c|c|}
 \hline
 \textbf{Algorithms}  & \textbf{Planning phase} & \textbf{Transmission phase}\\ 
 \hline
  FHJPA &  $\mathcal{O}((K-1){N_S}^2N_A)$ & $\mathcal{O}(K)$\\ 
  \hline
  GA & $-$ & $\mathcal{O}(KN_A)$\\
  \hline
  IHJPA &  $\mathcal{O}\big({N_A^{N_S}}/{N_S}\big)$ & $\mathcal{O}(K)$\\ 
  \hline
 \end{tabular}
 \caption{Complexities of different algorithms.}
  \label{complexity}
  \vspace{-0.75cm}
\end{table}

The computational complexities of the algorithms are compared in Table \ref{complexity} for both the planning and transmission phases.
The complexities of the FHJPA, GA, and IHJPA algorithms can be obtained following \cite{octavia2020VTC} and \cite{shalini_TVT_arxiv}, respectively. The GA does not involve a planning phase as it directly allocates power in the transmission phase without requiring a look-up table.

%%%%%%%%%%%%%%%%%%%%%%%%%%%%%%%%%%%%%%%%%%%%%%%%%%%%%%%%%%%%%%%%%%%%%%%%%%%%%%%%%%%%%%%%%%%%%%%%%%%%%%
\section{Conclusion}
% \vspace{-.1cm}
This work investigates a joint allocation of transmit power at the source and jamming power at the destination to maximize the average secrecy energy efficiency of a system equipped with energy harvesting devices with limited battery capacities over a finite time slot. 
% As our system follows the Markov property, we formulate an online sequential decision-making problem using a finite-horizon MDP framework. 
We propose a finite-horizon algorithm (FHJPA) and compare it with a low-complexity greedy algorithm (GA). Additionally, we propose an infinite-horizon algorithm (IHJPA) to compare it with the finite-horizon algorithm. We observe that the FHJPA algorithm outperforms other algorithms due to the accuracy of modelling the problem using finite-horizon MDP. As the time horizon increases, the accuracy of the infinite-horizon modelling improves and the performance gap between FHJPA and IHJPA algorithms reduces. We also find that greedily optimizing the immediate reward in each time slot, a low-complexity technique, may be nearly optimal when sufficient energy is available in the source's battery. 
% We observe that the performance of gradually converges toward that of the FHJPA algorithm. 
 The computational complexities of all algorithms are also analyzed. The FHJPA algorithm is found to take $16.6$ percent less computational time compared to the IHJPA algorithm.  We finally conclude that system optimization should strike a balance between maximizing average secrecy energy efficiency and increasing the total number of securely transmitted bits rather than optimizing a single performance metric.

%%%%%%%%%%%%%%%%%%%%%%%%%%%%%%%%%%%%%%%%%%%%%%%%%%%%%%%%%%%%%%%%%%%%%%%%%%%%%%%%%%%%%%%%%%%%%%%%%%%%%%
% \vspace{-.25cm}

\section*{Acknowledgement}
% \vspace{-.2cm}
This work was supported in part by the Science and Engineering Research Board (SERB), India, under the Core Research Grant CRG/2023/004869 and by Taighde Éireann – Research Ireland under Grant number 22/PATH-S/10788.
 
% \vspace{-.1cm}
\bibliographystyle{IEEEtran}

\bibliography{IEEEabrv,PA_RL}

\end{document}